\documentclass[times,review]{elsarticle}
\usepackage{amssymb,amsmath}
\usepackage{graphicx}
\usepackage{natbib}
\usepackage{multirow}
\usepackage{amsfonts}
\usepackage{siunitx}
\usepackage{textcomp}
\usepackage[dvipsnames]{xcolor}
\usepackage[utf8]{inputenc}
\usepackage{mhchem}
\usepackage{lineno,hyperref}
\modulolinenumbers[5]
\biboptions{numbers,sort&compress}
\begin{document}
\begin{frontmatter}
\title{Ultrahigh nitrogen-vacancy center concentration in diamond}

\author[add1]{S. Kollarics}
\author[add1,add2,add3]{F. Simon\corref{corresp}}
\ead{simon.ferenc@ttk.bme.hu}
\author[add1]{A. Bojtor}
\author[add1]{K. Koltai}
\author[add4]{G. Klujber}
\author[add4]{M. Szieberth}
\author[add1,add3,add10]{B. G. M\'{a}rkus}
\author[add3,add5]{D. Beke}
\author[add3]{K. Kamar\'{a}s}
\author[add3,add5]{A. Gali}
\author[add6]{D. Amirari}
\author[add6]{R. Berry}
\author[add6]{S. Boucher}
\author[add6]{D. Gavryushkin}
\author[add7]{G. Jeschke}
\author[add8]{J. P. Cleveland}
\author[add9]{S. Takahashi}
\author[add2]{P. Szirmai}
\author[add2,add10]{L. Forr\'{o}}
\author[add11]{E. Emmanouilidou}
\author[add11]{R. Singh}
\author[add11]{K. Holczer}

\cortext[corresp]{Corresponding author}

\address[add1]{Department of Physics, Budapest University of Technology and Economics and MTA-BME Lend\"{u}let Spintronics Research Group (PROSPIN), P.O. Box 91, H-1521 Budapest, Hungary}
\address[add2]{Laboratory of Physics of Complex Matter, \'{E}cole Polytechnique F\'{e}d\'{e}rale de Lausanne, Lausanne CH-1015, Switzerland}
\address[add3]{Institute for Solid State Physics and Optics, Wigner Research Centre for Physics, PO. Box 49, H-1525, Hungary}
\address[add4]{Institute of Nuclear Techniques, Budapest University of Technology and Economics, M\H{u}egyetem rkp. 9, H-1111 Budapest, Hungary}
\address[add5]{Department of Atomic Physics,  Budapest University of Technology and Economics, Budafoki út 8., H-1111 Budapest, Hungary}
\address[add6]{RadiaBeam, 1735 Stewart Street, Suite A, Santa Monica, California 90404}
\address[add7]{Department of Chemistry and Applied Biosciences, Swiss Federal Institute of Technology, Z\"urich, Switzerland}
\address[add8]{SomaLogic, Inc. 2945 Wilderness Pl, Boulder, Colorado 80301, USA}
\address[add9]{Department of Chemistry, University of Southern California, Los Angeles CA 90089, USA}
\address[add10]{Stavropoulos Center for Complex Quantum Matter, Department of Physics, University of Notre Dame, Notre Dame, Indiana 46556, USA}
\address[add11]{Department of Physics, University of California at Los Angeles, Los Angeles, California 90024, USA}

\begin{keyword}
	nitrogen-vacancy center ensembles, neutron irradiation, electron irradiation, electron paramagnetic resonance, relaxation time, electron-nuclear double resonance
\end{keyword}

\begin{abstract}
High concentration of negatively charged nitrogen-vacancy ($\text{NV}^{-}$) centers was created in diamond single crystals containing approximately 100 ppm nitrogen using electron and neutron irradiation and subsequent thermal annealing in a stepwise manner. Continuous-wave electron paramagnetic resonance (EPR) was used to determine the transformation efficiency from isolated N atoms to $\text{NV}^{-}$ centers in each production step and its highest value was as high as 17.5\%. Charged vacancies are formed after electron irradiation as shown by EPR spectra, but the thermal annealing restores the sample quality as the defect signal diminishes. We find that about 25\% of the vacancies form NVs during the annealing process. The large $\text{NV}^{-}$ concentration allows observing orientation dependent spin-relaxation times and also determining the hyperfine and quadrupole coupling constants with high precision using electron spin echo (ESE) and electron-nuclear double resonance (ENDOR). We also observed the EPR signal associated with the so-called W16 centers, whose spectroscopic properties might imply a nitrogen dimer-vacancy center for its origin.  
\end{abstract}

\end{frontmatter}
%\twocolumn

%\linenumbers
\section{Introduction}
Impurities in solid state materials have been in the focus of attention for many decades due to their interesting and complex physics and promising applications. The negatively charged nitrogen-vacancy ($\text{NV}^{-}$) centers \cite{Clark1956,Loubser1978} are indisputably among the most studied ones. Individual $\text{NV}^{-}$ centers \cite{Gruber1997} can be used as single-photon sources \cite{Kurtseifer2000singlephoton}, solid state spin quantum bits \cite{Jelezko2004qubit} and they have enormous potential in nanoscale metrology e.g. in magnetometry \cite{Maze2008magnetometry}, in electrometry \cite{Chen2017electrometry}, or in thermometry \cite{Neumann2013thermo}. The biocompatibility of diamond and the nanometer scale spatial resolution of NV-based thermometers makes them appealing also for life-science applications \cite{Kucsko2013biolthermo} as temperature plays a crucial role in biological processes. Nanodiamonds containing $\text{NV}^{-}$ centers \cite{Kumar2019cvd} are non-invasive tools for studying the magnetism of biological samples such as proteins \cite{Ermakova2013biol}.

Using ensembles of $\text{NV}^{-}$ centers improves the magnetic field sensitivity of magnetic imaging by a factor of $\sqrt{N}$, where $N$ is the number of $\text{NV}^{-}$ centers in the focus spot \cite{Rondin2014magnetoReview}. The detection limit of high density ensembles of $\text{NV}^{-}$ centers in diamond was proved to lie in the $\text{fT}/\sqrt{\text{Hz}}$ range and even lower limits could be reached \cite{Taylor2008}. The spatial resolution of such systems is diffraction limited but this can be overcome by stimulated emission depletion fluorescence microscopy (STED) \cite{Westphal2005STED,Rittweger2009STED}. However, to create such sensitive magnetometers, competing with the present best performing techniques, it is important to have control over the production of $\text{NV}^{-}$ centers and have a reliable technique to quantitatively classify the prepared samples. The study of $\text{NV}^{-}$ ensembles allows the use of macroscopic experimental methods which are not feasible on individual $\text{NV}^{-}$ centers and also a simultaneous investigation of $\text{NV}^{-}$ units along different crystallographic directions, thus allowing for a comparative study of orientation dependent physical properties. The realization of dense spin ensembles enables to study long-range interactions and many-body localization. This was presented on high density NV center ensembles in Ref. \cite{Choi2017DTC}.

The NV center in diamond is a complex defect consisting of a vacancy and a nearest neighbor single substitutional nitrogen atom. The resulting 5 electron system is called the neutral NV center ($\text{NV}^0$) and possesses a $\text{C}_{3v}$ symmetry. This center can catch an additional electron from a second nitrogen in its vicinity creating positively charged nitrogen atoms in the diamond lattice \cite{Lawson1998posnitroftir}. This negatively charged NV center ($\text{NV}^{-}$) has spin $\text{S}=1$ and its electronic ground (and excited) state is a triplet enabling the manipulation through microwaves. The unique electronic structure of the $\text{NV}^{-}$ center and the excellent physical properties of diamond are the reasons behind the above-listed versatile application ideas. The spin Hamiltonian of the electronic ground state includes the electronic Zeeman interaction, zero-field splitting, nuclear Zeeman interaction, anisotropic hyperfine, and axial nuclear quadrupole terms:
\setlength{\belowdisplayskip}{6pt} \setlength{\belowdisplayshortskip}{0pt}
\setlength{\abovedisplayskip}{6pt} \setlength{\abovedisplayshortskip}{0pt}
\begin{equation*}
	\begin{split}
		H=\mu_B B g_e S+D\left(S_z^2-\frac{1}{3}S\left(S+1\right)\right)-\mu_n B g_n I+ \\
		+A_\perp\left(S_xI_x+S_yI_y\right)+ A_\parallel S_zI_z + P\left(I_z^2-\frac{1}{3}I\left(I+1\right)\right)
	\end{split}
\end{equation*}

Here, we study the production of macroscopic amounts of $\text{NV}^{-}$ centers from commercially available diamonds using a systematic, stepwise electron and neutron irradiation and a subsequent annealing process. We show that the $\text{NV}^{-}$ production can be well followed with continuous-wave electron paramagnetic resonance (EPR). The conversion efficiency from isolated nitrogen atoms to $\text{NV}^{-}$ centers is as high as 17.5\% according to EPR. Electron irradiation produces charged vacancies with $S=1/2$, whose EPR signal is transformed into the $\text{NV}^{-}$ centers with an efficiency of about 25\% upon thermal annealing. The disappearance of the vacancy signal upon the annealing proves that the integrity of the diamond lattice is restored after each step. We also observe the EPR signal of the so-called W16 center in diamond which we assign to N-dimer-vacancy centers (NNV). Pulsed EPR on the $\text{NV}^{-}$ centers along different crystallographic directions reveals an orientation dependence of the spin-relaxation lifetimes. Electron-nuclear double resonance allows the precise determination of the nuclear hyperfine and quadrupole coupling constants.

\section{Experimental}
\subsection{EPR setup}
Samples produced by neutron irradiation were characterized using a Bruker Elexsys E500 continuous-wave (CW) spectrometer equipped with an ER 4122SHQE resonator. Diamonds were fixed to a quartz rod by using vacuum grease and the rod was inserted into a goniometer attached to the EPR cavity. Quantitative analysis of electron irradiated samples and pulse electron paramagnetic resonance studies were carried out on a Bruker Elexsys E580 FT/CW electron paramagnetic resonance spectrometer with a Flexline resonator (ER 4118X-MD5). A Bruker E580 spectrometer was used for the pulsed ENDOR measurements at X-band microwave frequency ($\sim$ 9.5 GHz). The Mims pulse sequence was used with the microwave and radiofrequency pulse lengths of $\tau_{\pi}$ = 16 ns and $\tau_{\mathrm{RF}}$ = 10 $\mu$s, respectively.

\subsection{Sample preparation}
Single crystal, type 1b diamond samples (supplied by Element Six Ltd.) produced by high pressure high temperature method (HPHT) containing less than $200\ \text{ppm}$ substitutional nitrogen were exposed to neutron irradiation at maximum power of $100\ \text{kW}$ for 8 hours in the Training reactor of the Institute of Nuclear Techniques at Budapest University of Technology and Economics. Three samples received a total neutron fluence of $10^{17}\ 1/\text{cm}^2$ ($3\cdot 10^{16}\ 1/\text{cm}^2$ in the $100\ \text{eV}$ - $1\ \text{MeV}$ range) which is in the low dose region \cite{Mita1996}. After that, they were annealed under dynamic vacuum ($10^{-6}\ \text{mbar}$) at $800\ \text{°C}$ based on the detailed annealing temperature studies found in Ref. \cite{Davies1992}. The irradiation and the subsequent annealing were repeated for two of the three samples.
Another six samples were exposed to electron irradiation using a variable-energy RF linear accelerator built by RadiaBeam Technologies LLC. The custom design incorporates a novel two accelerating structure system which produces an electron beam with variable energy between 1 and 4 MeV, up to 68 mA peak current and 68 µA average current. The samples received net fluence up to $2.8 \cdot 10^{18}\,1/\text{cm}^2$, which was determined from the resulting electric current. These samples were also annealed at $800\ \text{°C}$ for two hours and at $1000\,\text{°C}$ for another two hours. The irradiation and annealing steps were repeated several times resulting in a large set of $\text{NV}^{-}$ concentration data as a function of the electron fluence.

\section{Results and Discussion}
%\subsection{Growth of NV as followed in EPR}

\begin{figure}[!h]
	\centering
	\includegraphics*[width=\linewidth]{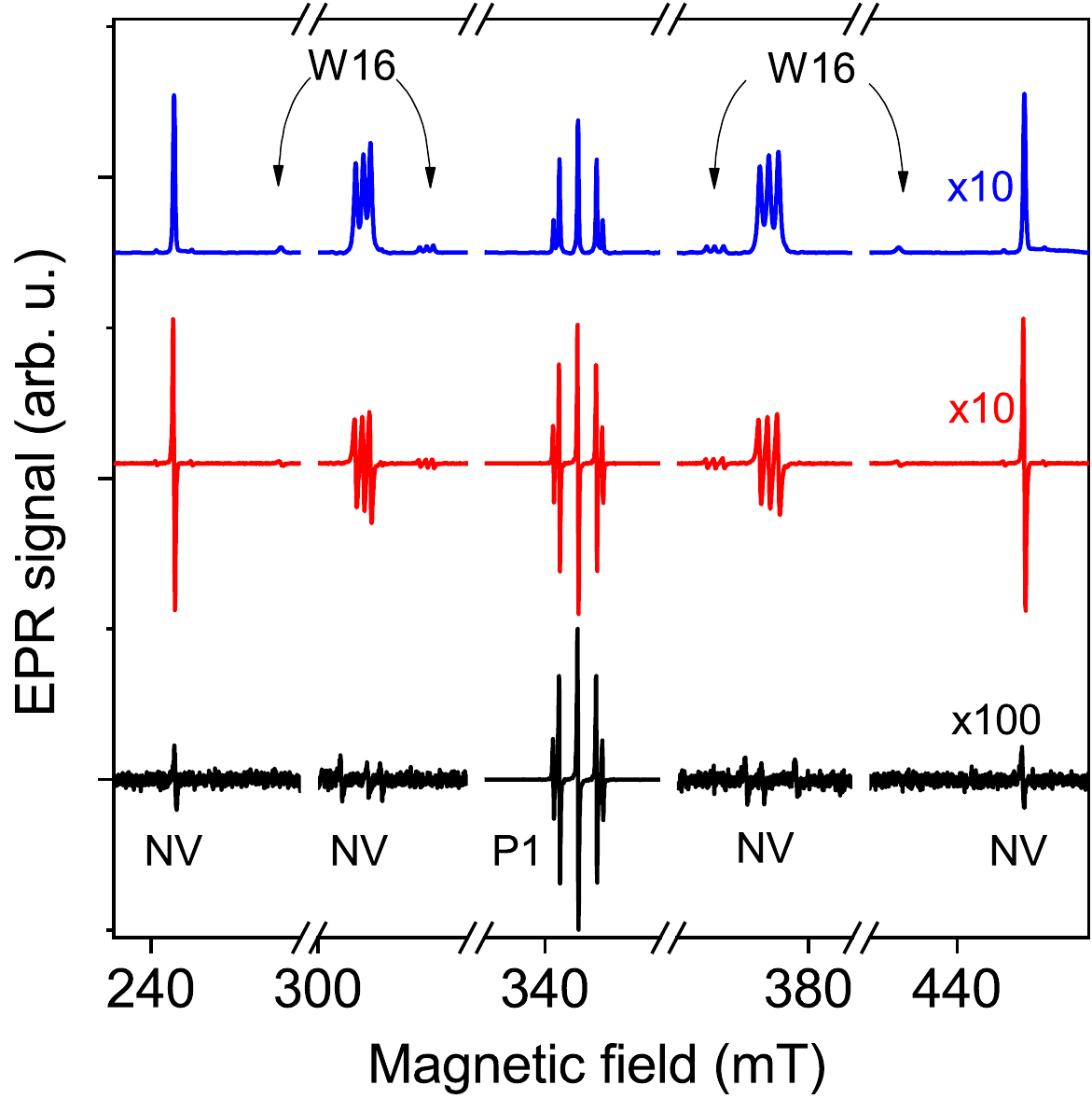}
	\caption{Electron paramagnetic resonance spectra acquired after the first (black line) and fifth (red line) electron irradiation and subsequent annealing step of sample S1. The spectra are normalized to the central nitrogen line also known as the P1 center. The spectral range, which contains the EPR signal of the $\text{NV}^{-}$ centers, is scaled up for clarity. The blue line shows the integrated EPR signal after the fifth irradiation-annealing step. Arrows point to the EPR signal of the W16 center.}
	\label{s1hex}
\end{figure}

Fig. \ref{s1hex} shows EPR spectra recorded after the first and fifth steps of electron irradiation followed by annealing. The single crystal sample was placed with its $<111>$ direction almost parallel with the external magnetic field. Substitutional nitrogen with $S=1/2$ (also known as the P1 center) as the main impurity gives the strongest signal in the middle of the spectrum. When the external magnetic field is parallel with the $<100>$ axis, three resonances with the same intensity can be observed due to the hyperfine splitting from the $I=1$ nuclear spin of $^{14}$N. By aligning the field in the $<111>$ direction, five resonances with the relative intensity ratio 1:3:4:3:1 appear indicating that the non-bonding electron sits on an antibonding p-type orbital \cite{Smith1959}. 

Although the starting material shows no sign of $\text{NV}^{-}$ centers even after an annealing cycle, the characteristic $\text{NV}^{-}$ lines emerge after the neutron or electron irradiation and subsequent annealing as shown in Fig. \ref{s1hex}. In a general alignment, eight lines related to $\text{NV}^{-}$ centers are present as there are four possible crystallographic orientations and two possible $\Delta m=1$ transitions. The EPR spectrum of the $\text{NV}^{-}$ is well understood \cite{Loubser1978}: the lower- and uppermost signal arises from the $0 \rightarrow 1$ and $-1 \rightarrow 0$ transition, respectively, for $\text{NV}^{-}$ centers that are oriented along the external field. The inner lying six lines correspond to NVs which are aligned at a $109.47$ degree angle with respect to the external field. 

Fig. \ref{s1hex} also shows the so-called W16 center according to the notation in Ref. \cite{Loubser1978}. This signal corresponds to a $S=1$ triplet spin state, similarly to the $\text{NV}^{-}$ center (which was denoted as W15 in Ref. \cite{Loubser1978}) but with a smaller zero-field splitting, \mbox{$D(\mathrm{W16})=0.86\cdot D(\mathrm{NV})$}, as shown in the figure. Given the similarity of this center to the NV, we tentatively assign it to N(+)NV(-) complexes. The EPR signal intensity of the W16 is about $9\%$ in the samples we investigated, a ratio which is independent of the thermal treatment and irradiation conditions. We therefore conclude that it originates from N-N dimers which are present in the starting samples. Interestingly, the W16 center does not show light-induced pumping with the 532 nm laser and it merits further studies. We underline that our samples allow for further studies of this potentially interesting, however yet elusive center in diamond.

We argue that EPR is the \emph{ideal} choice to determine the conversion efficiency from N to $\text{NV}^{-}$ as it spectroscopically differentiates between the centers and it is also sensitive to the number of spins. While it is generally more difficult to obtain absolute spin numbers, it is highly sensitive to relative changes in the number of different types of spins \cite{Atherton,Slichter}. 

\begin{figure}[!h]
	\centering
	\includegraphics*[width=\linewidth]{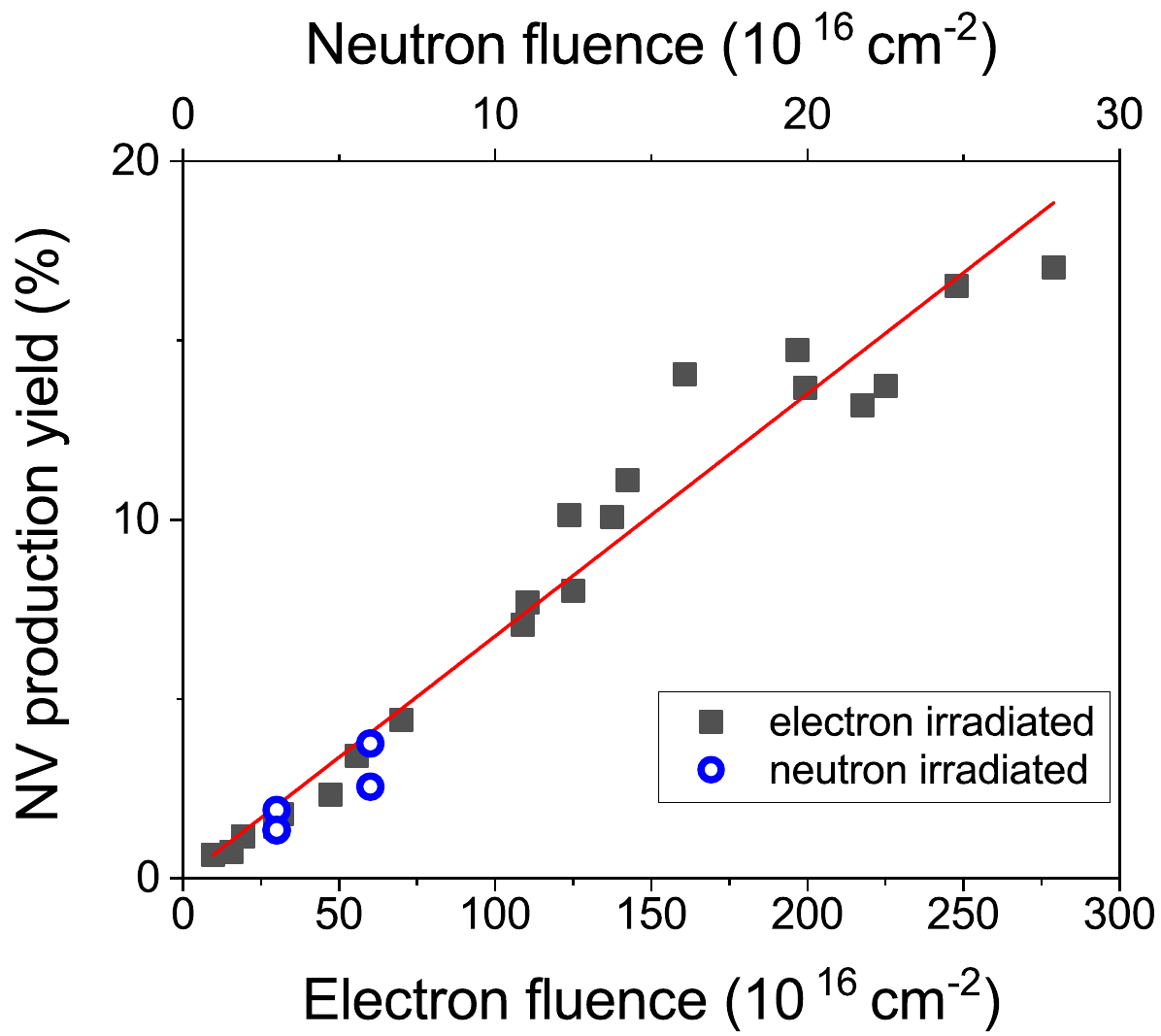}
	\caption{The production yield of $\text{NV}^{-}$ centers from individual nitrogens as obtained from the EPR signal for neutron irradiated (blue circles) and electron irradiated (black squares) samples as a function of the neutron end electron fluence. Note that the electron and neutron fluence ranges differ in a factor of 10.}
	\label{ratio}
\end{figure}

Fluorescence or optical absorption-based methods are powerful tools to qualitatively analyze color centers in diamond and in other solid materials. They are also widely used for quantitative measures \cite{Acosta2009}. However, these methods are more limited in determining the amount of color centers due to the inevitable and less center-specific optical absorption, which is significant in diamond. Studying neutron and electron irradiation, Nöbauer \emph{et al.} found that the type and energy of irradiation affect the optical transparency of the diamond host \cite{Nobauer2013arxiv}. Luminescence is also limited for quantitative purposes as assessment of the effective volume of detection and the ratio of backscattered photons is difficult due to the high refractive index of diamond. Suppression of photoluminescence through graphitization at high irradiation fluences was also reported \cite{Waldermann2007}, which also limits the possible use of this technique for quantitative analysis. Lattice distortions introduced by substitutional nitrogen in diamond also affect the PL intensity of $\text{NV}^{-}$ centers as it was shown that above $100\ \text{ppm}$ this results in an enormous decrease of PL signal \cite{Bogdanov2018}. Microwave radiation, on the other hand, penetrates into the sample, and the measured microwave absorption does not depend on the sample surface properties and nitrogen concentration has no such effect on signal intensity as described before. In addition, as mentioned we can use the concentration of the P1 center as a reference therefore even an altered penetration of microwaves would not affect the analysis.

Fig. \ref{ratio} shows the relative concentration of $\text{NV}^{-}$ centers as a function of the irradiation fluence. EPR spectra, like the ones presented in Fig. \ref{s1hex} were fitted with derivative Lorentzian lines to obtain the signal intensities. The sum of integrals related to $\text{NV}^{-}$ centers is compared to the sum of the integrals of $\text{NV}^{-}$ and nitrogen lines as shown in Fig. \ref{ratio}. In the analysis, one has to take into account the spin susceptibility difference between the P1 center ($S=1/2$) and the $\text{NV}^{-}$ ($S=1$), which goes as $S(S+1)$ therefore the integrated intensity of the $\text{NV}^{-}$ centers, $I_{\text{NV}}$ has to be multiplied by $q=3/8$ for the proper normalization, i.e. to obtain it as the equivalent number of $S=1/2$ spins. One has two options to obtain the $\text{NV}^{-}$ concentration or \emph{NV production yield}, $c_{\text{NV}}$ from the EPR data: 
$c_{\text{NV}}=\frac{q I_{\text{NV}}}{I_{\text{P1}}}$ or $c_{\text{NV}}=\frac{q I_{\text{NV}}}{\left(I_{\text{P1}}+q I_{\text{NV}}\right)}$. Although the resulting data is not significantly affected by this choice for moderate values of $c_{\text{NV}}$, we found the second choice to be more appropriate as it also reflects that the amount of the P1 centers drops during the P1$\rightarrow$NV transformation. 

Fig. \ref{ratio}. demonstrates that the experimentally determined $c_{\text{NV}}$ scales with the fluence for both electron and neutron irradiation, although there is a factor 10 difference depending on the type of irradiation. The NV yields suggest that neutrons cause an order of magnitude more damage compared to electrons which is consistent with previous simulations \cite{Mainwood1997neutrondamage,Campbell2000electrondamage}. For the highest $\text{NV}^{-}$ production yield, we observe $17.5\%$ in a sample with $70\ \text{ppm}$ nitrogen concentration and $15\ \text{ppm}$ $\text{NV}^{-}$ concentration, which are obtained from a careful calibration against a CuSO$_4\cdot 5$H$_2$O standard. This concentration clearly lags behind the $45~\text{ppm}$ of $\text{NV}^{-}$ centers reported in Ref. \cite{Kucsko45ppm2018} but considering that the total electron fluence reported in that work was $1.4 \cdot 10^{19}\,1/\text{cm}^2$ which is five times higher than our net fluence, the difference is understandable. We also note that the same sample was reported to have $15\ \text{ppm}$ of $\text{NV}^{-}$ centers \cite{Zhou15ppm2020} which questions the accuracy of pulsed techniques for spin density estimation. 

\begin{figure}[!h]
	\centering
	\includegraphics*[width=\linewidth]{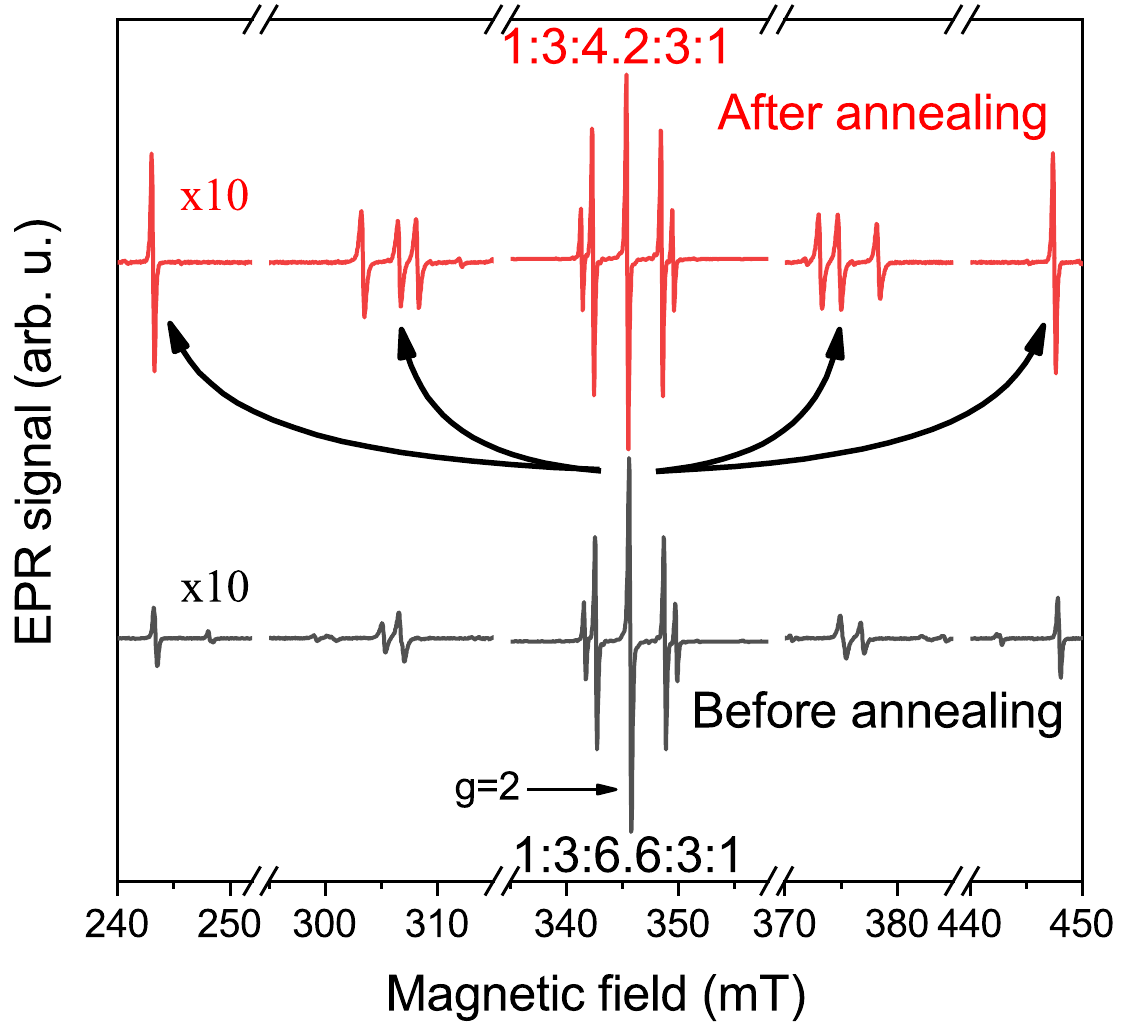}
	\caption{EPR spectra of an electron irradiated sample before and after the annealing step. Arrows indicate that the significant $g \approx 2$ line intensity is transformed into the $\text{NV}^{-}$ lines (the $\text{NV}^{-}$ spectral range is scaled up for better visibility by a factor of 10). The sample orientation is slightly different in the two spectra thus the $\text{NV}^{-}$ line positions are shifted but the line intensities are unaffected by the orientation. Note that the intensity of the central line has a small, about 2\%, additional intensity, which is present in the pristine samples, too. It is most probably due to some heteroatoms which are unaffected by the treatments.}
	\label{Before_After_annealing}
\end{figure}

Our stepwise $\text{NV}^{-}$ synthesis allows following the growth mechanism of the $\text{NV}^{-}$ center as well as the radiation damage to the sample. Prior to our work, it was unknown whether the annealing restores the diamond quality and what the conversion efficiency of $\text{NV}^{-}$ from the vacancies is.

Fig. \ref{Before_After_annealing} shows the EPR spectra of an electron irradiated sample before and after the annealing. The irradiation induces an additional signal at $g=2$, which overlaps with the middle line of the nitrogen EPR signal and thus appears as an additional intensity amounting to about 20\% of the total EPR intensity in the figure. We note that such a signal appears exclusively for the electron irradiated samples, most probably because it is due to a charged defect or V(-). However, neutron irradiation appears to produce uncharged defects only, which are EPR silent. The annealing transforms this additional intensity to the $\text{NV}^{-}$ lines (indicated by arrows) with a yield of about 25\%, i.e. according to EPR, every fourth EPR active defect forms an NV. When calculating the yield, i.e. counting the spins, we normalized the $\text{NV}^{-}$ signal intensity by 8/3 due to the higher spin susceptibility of $S=1$ spins, as mentioned earlier.

Following the annealing, we do not observe the expected 1:3:4:3:1 ratio of the nitrogen lines, and some additional $g=2$ signal is observed. However, such a signal is present in the as-received samples, i.e. without irradiation or annealing. Its intensity amounts to about 2\% of that of the individual nitrogens. This is probably due to about 1-2 ppm of additional heteroatoms, whose presence cannot be ruled out according to the diamond manufacturer. We therefore conclude that the annealing restores the integrity of the diamond lattice, even though the irradiation induces a sizable amount of defects.

%\subsection{Angular dependence of the spin-relaxation properties}

\begin{figure}[!h]
	\centering
	\includegraphics*[width=\linewidth]{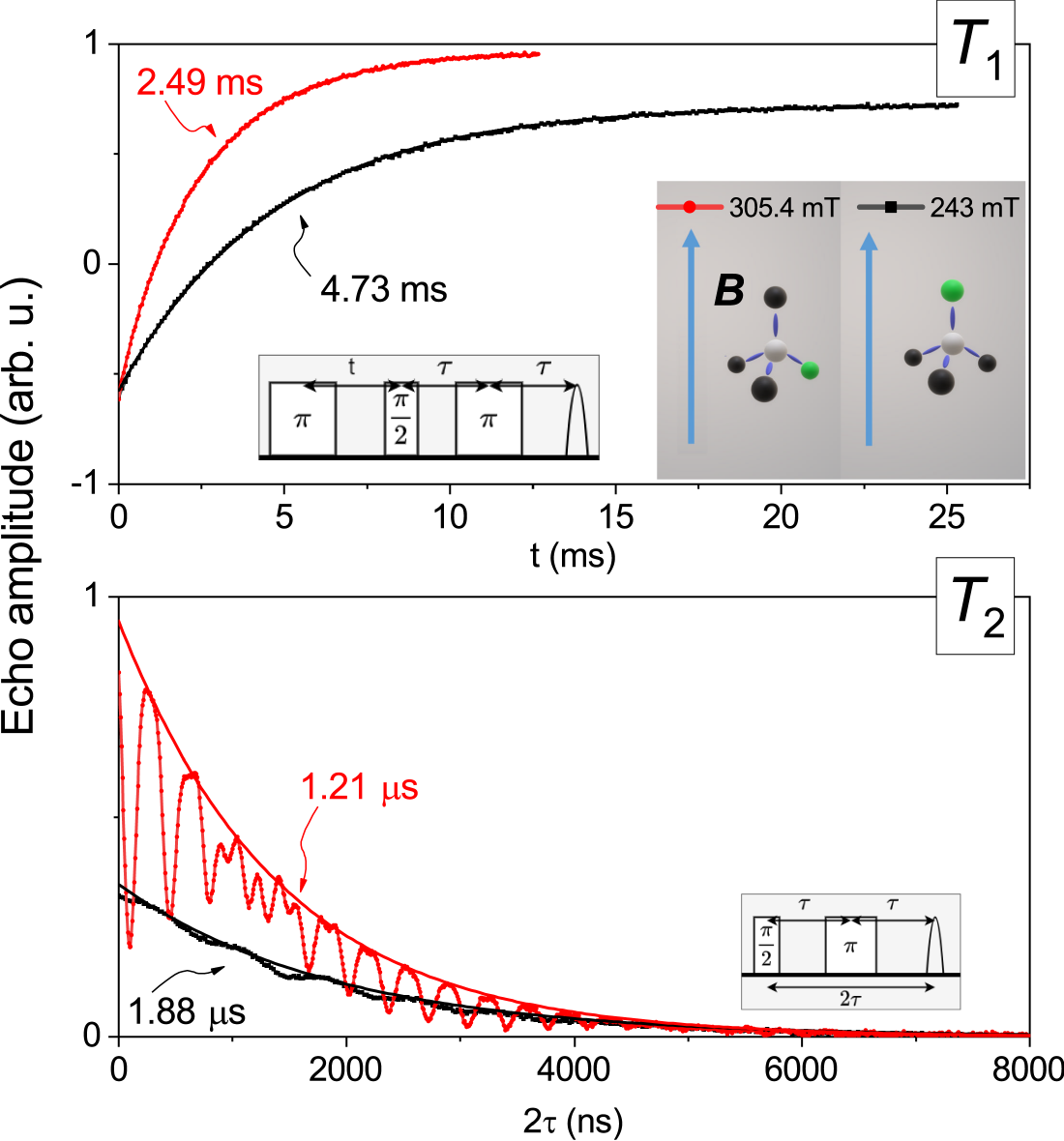}
	\caption{\emph{Angular dependent spin-lattice ($T_1$) and spin-spin relaxation times ($T_2$), measured with Hahn-echo detected inversion recovery and Hahn-echo decay, respectively. Note the different timescales for the two sets of data and the indicated relaxation time values. Insets show the applied pulse sequences and the orientation of NV centers with respect to the external magnetic field.}}
	\label{pulsed_ESR}
\end{figure}

The prospective applications of the $\text{NV}^{-}$ center as sensors \cite{Q_Sensors}, qubits \cite{Gruber1997}, or microwave sources \cite{MaserNature} rely heavily on the magnitude of the spin-lattice ($T_1$), spin-spin ($T_2$), and spin-decoherence ($T_2^*$) relaxation times. Optically detected magnetic resonance (ODMR) is in principle capable of measuring all three relaxation times \cite{Nivotsev2005,Pham2011}, but the information relies on measuring the populations of the individual spin levels rather than the collective Larmor precession of the spins in a spin echo experiment. In other words, ODMR detects the projection of the ensemble spin magnetization onto the $z$-axis of the Bloch sphere, while the conventional spin echo measures directly the collective Larmor precession of the spins. In addition, the optical methods are less representative for a bulk sample and they are also limited for samples with a high extinction coefficient. While these arguments are true for an ensemble sample (which is thought to be better suited for sensor applications) these parameters are hardly accessible without resorting to ODMR.

While ODMR studies of the spin relaxation times were done extensively with various concentrations of $\text{NV}^{-}$ ensembles, EPR investigations have been less explored due to the limited sensitivity of the measurement \cite{Shames, Mindarava2020, Rose2017, Takahashi2008t1t2}. Knowledge of $T_1$ is crucial for the sensor applications as well as it defines the longest timescale over which quantum information may be stored. Measurement of $T_2$ is also crucial for the timescale of coherent quantum information storage. It is also predicted \cite{MazeAnisTheor} that the $T_2$ is strongly anisotropic, which was observed using ODMR \cite{Stanwix2010} but to our knowledge, no similar conventional spin echo study exists which could confirm this observation.

Fig. \ref{pulsed_ESR} shows the $T_1$ relaxation time measured with Hahn-echo detected inversion recovery and the $T_2$ relaxation time acquired by measuring the Hahn-echo decay. The advantage of the ensemble sample studies is that they allow for the measurement of these parameters for different alignments of the NV axis with respect to the external magnetic field. To our surprise, we observe a clear and reproducible anisotropy as both relaxation times are shorter for $\text{NV}^{-}$ centers which form a $109.47$ degree angle with respect to the external field in comparison to those whose axis is along the magnetic field. 

The most plausible origin of the anisotropy lies in the spin Hamiltonian: crystalline vibrations modulate the dominant zero-field splitting parameter (the $D$) which acts as a fluctuating field along the NV axis. It is well known \cite{Slichter,Abragambook} that fluctuating fields aligned perpendicular to the external DC magnetic field can shorten the relaxation times. This therefore explains why we observe shorter relaxation times when the NV axis is not parallel to the external field. Based on a simple derivation (see Supplementary Material:5), we expect the $T_1$ of $\text{NV}^{-}$ centers parallel to the external magnetic field to be 2.77 times longer than in the case when $\text{NV}^{-}$ centers are aligned in $109.47$ degree angle with respect to the external field. Experimentally we find this ratio to be $\approx1.9$, which indicates that there are other contributions to longitudinal relaxation such as the spin flips of $\ce{^{13}C}$ and other paramagnetic nuclei.

In principle, the magnitude of $T_2$ can serve as a good indicator of the NV content without the need for calibration against a spin standard. The dipole-dipole interaction between the NV spin leads to the effect of spin-diffusion, which results in a shortened $T_2$. We discuss this effect after Ref. \cite{Abragambook} in the Supplementary Information using the van Vleck formula. This quantitatively describes the magnitude of $T_2$ as a function of the average spin-spin separation. The calculation yielded a $T_2$ which is consistent with the NV content. In addition, Ref. \cite{Bauch2020} provides an empirical scaling of the observed $T_2$ of the NV spins with the nitrogen content. Using this result, our typical $T_2$ data of about 1.5 µs is consistent with the about 100 ppm nitrogen content.

Besides the overall exponential decay of the Hahn-echo, a modulation is visible in Fig. \ref{pulsed_ESR}, known as electron spin echo envelope modulation (ESEEM) \cite{Schweiger2001Jeschke, Mims_Nassau_McGee1961}. It is due to the hyperfine coupling of the electronic spin of NV centers with the $\ce{^{13}C}$ and $\ce{^{14}N}$ nuclei. The black curve in Fig. \ref{pulsed_ESR} shows the modulation caused by distant $\ce{^{13}C}$ nuclei with a nuclear Zeeman frequency of $2.56\ \text{MHz}$ (at $B=243\ \text{mT}$) which is in good agreement with our ENDOR measurements. In this orientation, the modulation from $\ce{^{14}N}$ coupling is zero as the principal axis of the hyperfine and quadrupole tensors is parallel to the external magnetic field, but strongly present in the other orientation (red curve in Fig. \ref{pulsed_ESR}) when these tensors have a perpendicular component. It is clear from the envelope modulation that it is not governed by a single frequency. In general, the envelope modulation frequencies are linear combinations of hyperfine and quadrupole couplings, and angular dependence should also be taken into account. A more detailed analysis of the observed signal (based on a larger data set and supporting simulation) is beyond the scope of this paper.

%\subsection{X-band ENDOR results on the NV ensemble samples}

A range of compelling possibilities for the application of the $\text{NV}^{-}$ centers is where the physics of the electron spin is combined with nearby nuclei. This can be exploited either in nucleus-based quantum computing \cite{MehringPRL2003,MehringPRL2004} or in optically detected NMR applications \cite{AwschalomNMR_SCI,WrachtrupNMR_SCI}. Clearly, a prerequisite for such investigation is the accurate knowledge of the electron-nuclear hyperfine couplings which is accessible from \mbox{ESEEM} measurements (with the above-mentioned difficulties) or from electron-nuclear double resonance (ENDOR) \cite{Schweiger2001Jeschke} experiments. 
	
The ENDOR investigations rely on detecting a change in the electron spin populations (or polarization) when a nuclear transition is excited; a change in the nuclear population affects the electron populations as the two subsystems are connected by the hyperfine interaction. Given that the nuclear transitions are usually very well defined (down to a few 100 Hz to a few kHz), the magnitude of the hyperfine interaction can be determined with high accuracy.

\begin{figure}[!h]
	\centering
	\includegraphics*[width=\linewidth]{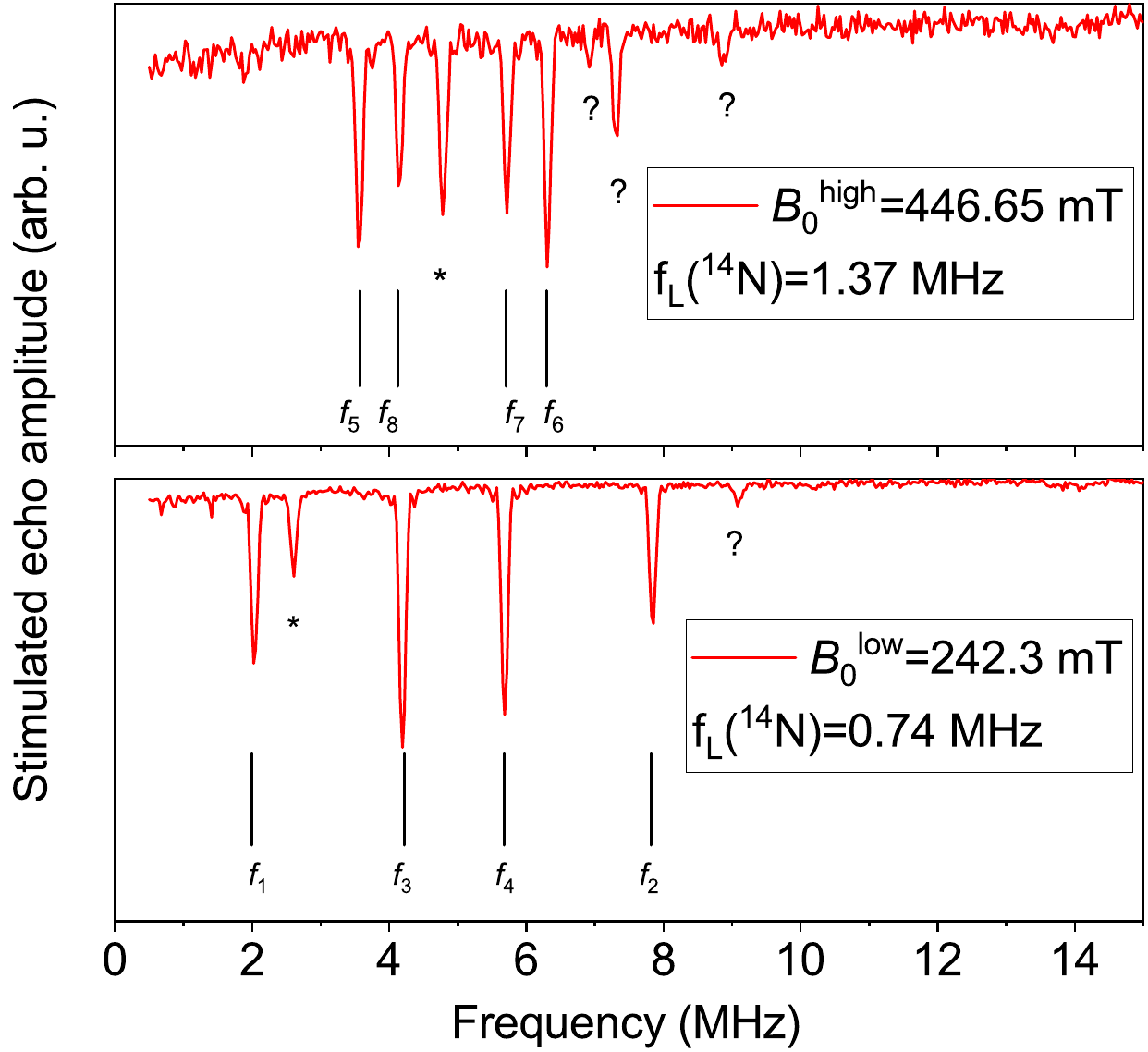}
	\caption{X-band ENDOR spectrum on a diamond $\text{NV}^{-}$ sample. The asterisks denote
		resonance corresponding to the nuclear Zeeman frequency of $\ce{^{13}C}$. The ENDOR frequencies using the notation of Ref. \cite{Yavkin2015} are given. ENDOR transitions denoted by question marks could not be unambiguously assigned.}
	\label{endor}
\end{figure}

ENDOR data using the Mims-ENDOR pulse sequence \cite{MimsEndor1965} on one of our samples is presented in Fig. \ref{endor}. for the highest and lowest lying $\text{NV}^{-}$ signals ($B_0^{\text{high}}=446.65\ \text{mT}$ and $B_0^{\text{low}}=242.3\ \text{mT}$) when the sample is oriented with its $<111>$ axis parallel to the field. 

To our knowledge, ours is the first ENDOR observation in low-field/frequency (at 9.5 GHz) besides an earlier report at 94 GHz \cite{Yavkin2015}. The lack of any publicly available X-band ENDOR data on the $\text{NV}^{-}$ center is probably a result of the somewhat complicated line assignment in low magnetic fields as shown below, and also that the ENDOR signal intensity scales with a higher power of the respective Larmor frequencies.

Our quantitative analysis of the observed ENDOR lines builds on the high-field ENDOR data  \cite{Yavkin2015} and on the hyperfine data obtained from conventional EPR experiments \cite{Smeltzer2009} and the frequencies of the major lines can be assigned as follows:

%\begin{equation}
\begin{align*}
f_1&=\left| f_{\text{L}}^{\text{low}}+P-A\right|\\
f_2&=f_{\text{L}}^{\text{low}}-P-A\\
f_3&=\left| f_{\text{L}}^{\text{low}}+P\right|\\
f_4&=f_{\text{L}}^{\text{low}}-P\\
f_5&=\left| f_{\text{L}}^{\text{high}}+P\right|\\
f_6&=f_{\text{L}}^{\text{high}}-P\\
f_7&=\left| f_{\text{L}}^{\text{high}}+P+A\right|\\
f_8&=f_{\text{L}}^{\text{high}}-P+A\\
\end{align*}
%\end{equation}

where we used the $f_{1\dots8}$ notations introduced in Ref. \cite{Yavkin2015} and $f_{\text{L}}^{\text{high}}=1.37\ \text{MHz}$ and $f_{\text{L}}^{\text{low}}=0.74\ \text{MHz}$ denote the $^{14}$N Larmor frequency for the high and low magnetic fields, respectively. The ENDOR frequencies are due to the quadrupole interaction characterized by $P=-4.8~\text{MHz}$ and the hyperfine coupling of $A=-2.2~\text{MHz}$ for the given magnetic field orientation. The $\left|\dots\right|$ notation reflects that the ENDOR transitions are observable for both stimulated emission and absorption.

The asterisk at $2.605\pm0.015\ \text{MHz}$ in the low field spectrum and at $4.778\pm0.015~\text{MHz}$ in the high field one denotes the nuclear magnetic resonance of $\ce{^{13}C}$ which is in good agreement with the literature value of $\gamma_n/2\pi=10.7084~\text{MHz/T}$ considering the accuracy of the measurement of the external magnetic field.
 
The present ENDOR study is enabled by the high $\text{NV}^{-}$ concentrations in our samples; these can thus be useful for future magnetic resonance experiments in e.g. sensor applications such as sensing the vector of magnetic fields \cite{Soshenko2021gyroscope}. Besides, given that ENDOR is a potential candidate for the realization of solid-state quantum computing \cite{MehringPRL2003,MehringPRL2004}, our study opens the way for this realization in moderate magnetic fields and frequencies.

\section{Conclusion}
We presented the stepwise synthesis and characterization of nitrogen-vacancy centers in neutron/electron irradiated HPHT diamond. Based on CW electron paramagnetic resonance, we quantitatively analyzed the samples and determined the $\text{NV}^{-}$ center concentrations. We found that the EPR based sample characterization can be best used for samples with high density of $\text{NV}^{-}$ centers where optical methods fail. The highest attained $\text{NV}^{-}$ concentration is 15 ppm. We observe the signal of charged vacancies in electron irradiated samples which disappears upon thermal annealing and the defects transform to $\text{NV}^{-}$ with a 25\% transformation efficiency. After each irradiation and annealing step, a high-quality diamond sample is recovered thus a lasting irradiation has no permanent effects. Pulse electron paramagnetic resonance reveals that spin-lattice and spin-spin relaxation times depend on the $\text{NV}^{-}$ center orientation. We also presented the first X-band ENDOR results on the diamond $\text{NV}^{-}$ and the hyperfine and quadrupole couplings determined by carrying out electron-nuclear double resonance experiments are in good agreement with previous studies.

\section{Acknowledgment}
A. J\'anossy is acknowledged for enlightening discussions. This work was supported by the Hungarian National Research, Development and Innovation Office (NKFIH) Grant nos. K137852, FK 125063 and 2017-1.2.1-NKP-2017-00001. The work was supported by the Quantum Information National Laboratory sponsored by the Ministry of Innovation and Technology via NKFIH. A. Gali acknowledges the support from National Excellence Program (NKFIH Grant no. 129866), the European Comission for the project Asteriqs (Grant No. 820394) and the EU QuantERA for the project Q-Magine (NKFIH Grant No. 127889). S. Takahashi thanks support from the National Science Foundation (NSF CHE-2004252 with partial co-funding from the Quantum Information Science program in the Division of Physics). K. Holczer and J. Cleveland acknowledge support from the National Science Foundation, Convergence Accelerator 2040520 program.

%\section{*References}
\bibliography{nv_2020_resubmit}

\clearpage
\begin{center}
	\Large{Supplementary material to: Ultrahigh nitrogen-vacancy center concentration in diamond}
\end{center}
\normalsize

\begin{abstract}
This supplementary material is organized as follows. We give a more detailed description of the linear accelerator used for electron irradiation built by \mbox{RadiaBeam} Technologies LLC. We present a set of EPR spectra recorded after each step of irradiation and subsequent annealing on sample S1. After that, we give an estimation of $T_2$ based on dipole-dipole interactions between NV centers. In the main text, we presented spin-lattice and spin-spin relaxation times acquired by pulsed EPR techniques. In this supporting material, we include our results on the saturation of CW EPR spectrum supporting the angular dependence observed in pulsed EPR measurements. In the main text, we motivated this observation based on the angular dependence of ZFS induced relaxation. Here, we give the full derivation of this model. In the following section, integrated CW EPR spectrum is compared to a spectrum acquired by pulsed techniques, namely to the field swept Hahn-echo decay. To support our interpretation of the measured ESEEM signal, the Fourier transform of the \mbox{ESEEM} signal in the parallel case is included here. Finally, to make our \mbox{ENDOR} spectra more transparent, we summarize the resonance frequencies shown in Fig. 5 in the main text. 
\end{abstract}

\linenumbers
\section{Electron irradiation apparatus}
For the electron irradiation, we used a custom, variable-energy RF linear accelerator (LINAC) built by RadiaBeam Technologies, LLC. The custom design incorporates a novel two accelerating structure system which produces an electron beam with variable energy between 1 and 4 MeV, up to 68 mA peak current and 68 µA average current. Electron beam energy variation is achieved by incorporating a fixed 2.5 MeV accelerating structure and subsequently adding a second structure with variable phase and amplitude. By varying the phase and amplitude of the second structure, the fixed 2.5 MeV energy is decelerated to 1 or 2 MeV and similarly is accelerated to 3 or 4 MeV.

\begin{figure}[!h]
\centering
\includegraphics*[width=0.5\linewidth]{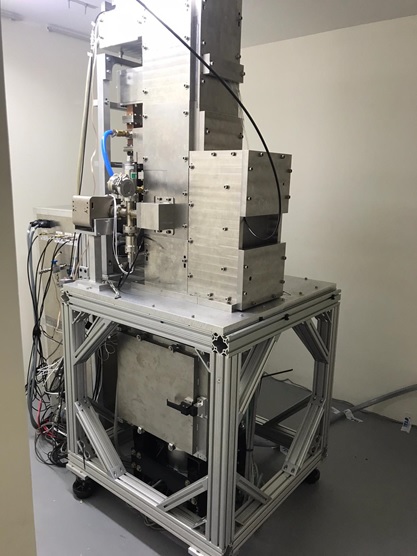}
\caption{\emph{Photograph of custom-built RF linear accelerator.}}
\label{linac}
\end{figure}

The initial electron beam is produced by a 15 kV modulated anode electron gun and accelerated through the structures at the specified energy. It is subsequently transported through an air cooled titanium window into a chamber where items can be placed for irradiation. The chamber can be operated under rough vacuum, which is pulled through a conflat flange with a standard scroll pump. For this experiment the base vacuum attained was $10^{-4}\ \text{torr}$. The average current was measured using an isolated platform where the diamonds were placed. As a note, the chamber can also be used with gas pressure. Fig. \ref{linac} shows a photograph of the LINAC. 

\clearpage

\section{EPR spectra recorded after each step of irradiation on sample S1}
Sample S1 was irradiated with electrons and subsequently annealed in five steps. After each annealing, EPR spectra were acquired as shown in Fig. \ref{s1hex}. With increasing irradiation, the concentration of NV centers and W16 centers grows.

\begin{figure}[!h]
\centering
\includegraphics*[width=0.9\linewidth]{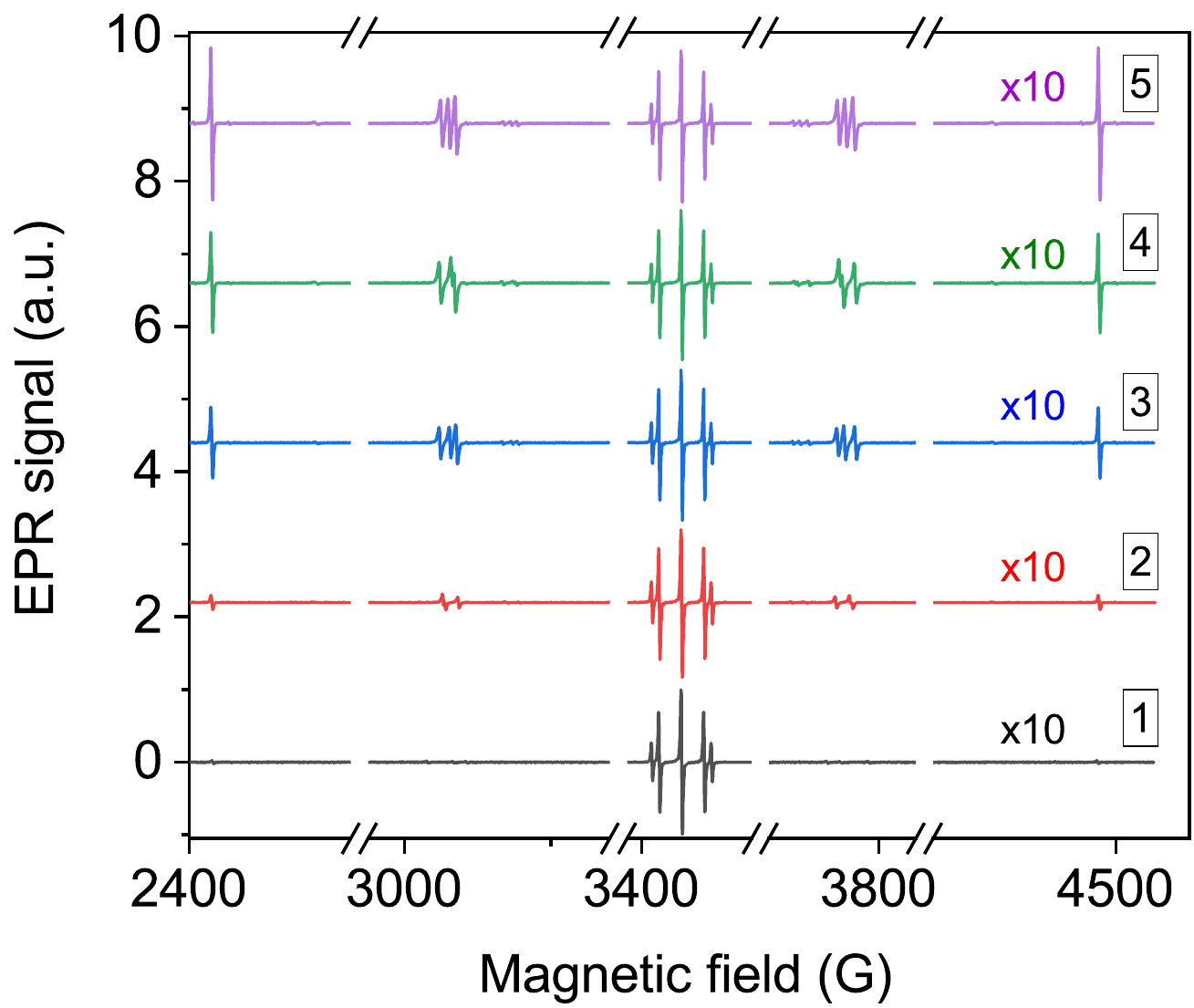}
\caption{\emph{Electron paramagnetic resonance spectra acquired after each irradiation and subsequent annealing of sample S1. The spectra are normalized to the central nitrogen line also known as the P1 center. Note the stepwise increase in NV center and W16 center concentration. For clarity, all the spectra are scaled up with a factor of 10 relative to their P1 lines.}}
\label{s1hex}
\end{figure}

\clearpage
\section{Estimate of $T_2$ from the van Vleck formula}
The experimental $T_2$ can be estimated from the van Vleck formula \cite{Abragambook}, which evaluates the dipole broadening. $T_2$ arises from the dipole-dipole interaction between like-spins and is not affected by the interaction with unlike spins, the magnetic field inhomogeneity, a distribution of the zero-field splitting due to e.g. strain, or an unresolved hyperfine coupling. The result reads:
\begin{equation} 
\frac{1}{T_2}=c\frac{\mu_0}{4\pi}\frac{g^2 \mu_{\text{B}}^2}{\hbar}\sqrt{\frac{3}{4}S(S+1)\sum_k\frac{\left(1-3\text{cos}^2\theta_{jk}\right)^2}{r^6_{jk}}},
\end{equation} 
\noindent where $\mu_0$ is the vacuum permeability, $r_{jk}$ is the length of the vector, $\mathbf{r_{jk}}$, from the origin (where the $j$th NV center is located) to the $k$th NV center, and $\theta_{jk}$ is the angle between the spin direction (it matches that of the external field for $B>0.1$ T) and $\mathbf{r_{jk}}$. 
The constant $c$ stands for the concentration as the dipolar linewidth or $1/T_2$ is linear with the concentration of the like-spins, due to the overall $1/r^3$ dependence in the formula.

A numerical evaluation of the sum yields for the diamond lattice $\sum_k..=$ 119.3, 746.4, and 955.4 when the magnetic field is along [100], [110], and [111] directions respectively. This gives for the $3.57\ \text{\AA}$ cubic lattice constant $T_2=0.3\ \mu\text{s}$ when the magnetic field is along the [111] direction and $c=10\,\text{ppm}$, which is fairly close to the experimentally observed value.
\clearpage
\section{Saturation of CW EPR}
\begin{figure}[!h]
\centering
\includegraphics*[width=0.8\linewidth]{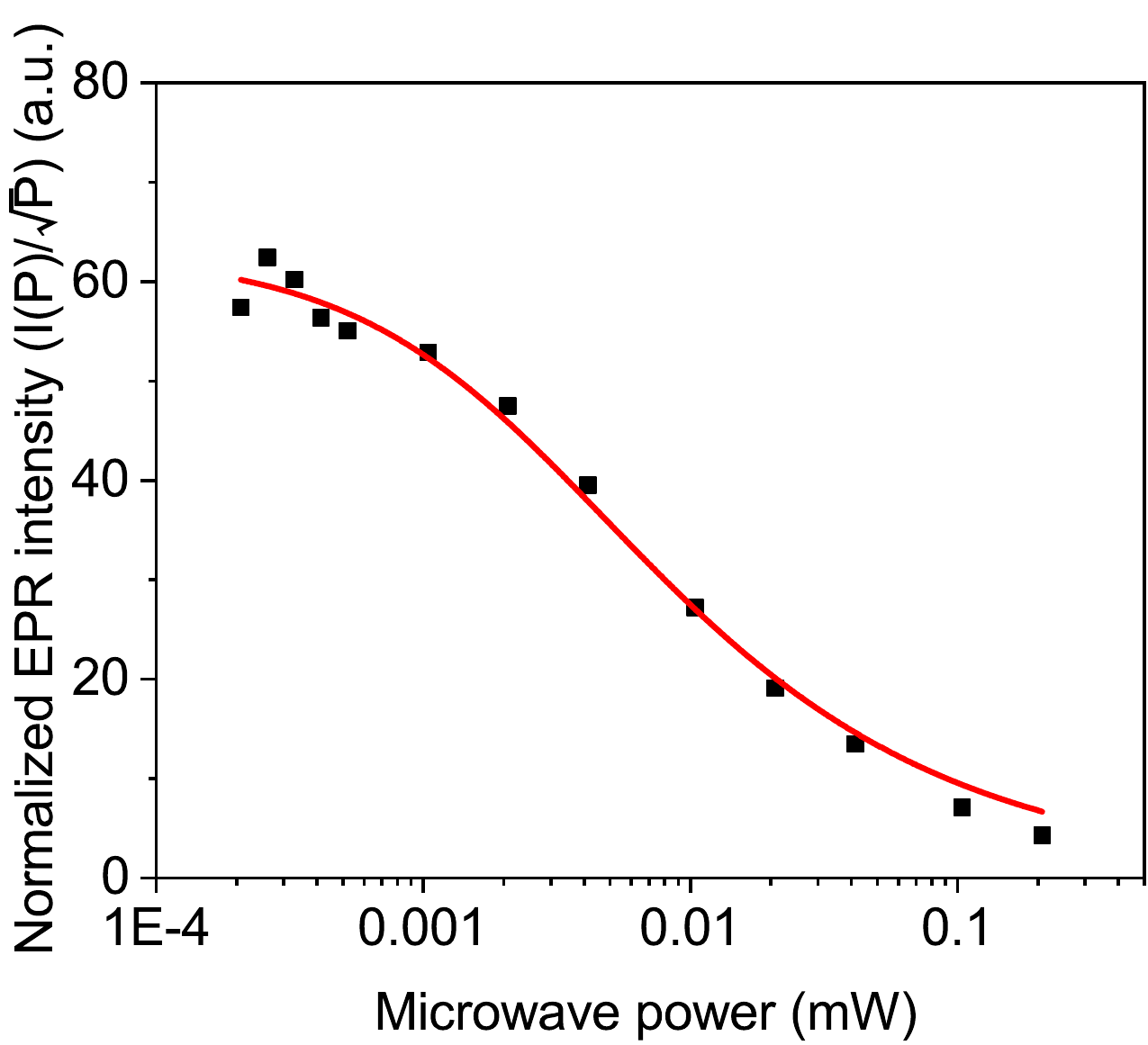}
\caption{\emph{CW EPR intensity of the lowest NV line. $T_1 T_2$ product is determined by fitting the saturation equation.}}
\label{cw_satur}
\end{figure}

In the main text, we presented pulse electron paramagnetic resonance measurements enabling us to determine the spin-lattice and spin-spin relaxation times of NV center ensembles. Even though this technique was tailored for such purposes, in certain cases there is a way to deduct from CW measurements to these values. The intensity of the CW EPR spectrum can be saturated by increasing the microwave power \cite{Portis1953}:

\begin{equation} 
\text{I(P)} \propto \frac{\sqrt{P}}{\sqrt{1+CQ\gamma^2 P T_1 T_2}},
\end{equation} 

\noindent where $P$ is the microwave power, $C$ is the cavity mode dependent conversion factor, $Q$ is the quality factor, and $\gamma$ is the electron gyromagnetic ratio often given in the following form: $\frac{\gamma}{2\pi}=28\ \frac{\text{GHz}}{\text{T}}$ . For our TE011 cavity $C=2.2\cdot 10^{12}\ \frac{\text{T}^2}{\text{W}}$ and the $Q$-factor was measured to be 5300 by the Bruker software. The product of the two relaxation times appear in the denominator while all the other factors can be measured therefore fitting the above equation one yields this product from microwave power dependent EPR intensity measurements such as shown in Fig. \ref{cw_satur} for one of the NV resonance lines.

\begin{table}[h!]
\begin{center}
\begin{tabular}{| c | c |}
	\hline
	Line & $T_1T_2 (10^{-12} \text{s}^2)$ \\
	\hline
	NV1 & $1164.5$ \\
	NV2 & $480.8$ \\
	NV3 & $323.7$ \\
	NV4 & $562.6$ \\
	NV5 & $505.2$ \\
	NV6 & $583.8$ \\
	NV7 & $457.6$ \\
	NV8 & $1194.7$ \\
	\hline
\end{tabular}

\end{center}
\caption{$T_1\cdot T_2$ calculated from the intensity of NV EPR lines in order of their resonant field (NV1-lowest to NV8-highest lying).}
\label{table:1}
\end{table}
The resulting relaxation time products listed in \ref{table:1} seem to resemble the orientation dependence of $T_1$ and $T_2$ directly measured by pulsed techniques presented in the main text. 
In CW EPR, $T_2$ determines the homogeneous linewidth of a spin packet, but in our case, the measured linewidth does not coincide with this. First of all, there is a weak hyperfine interaction which splits each line in three but the shift is only about $2.2\ \text{MHz}\approx 0.8\ \text{Gauss}$. The individual lines can be homogeneously broadened by the dipole-dipole interaction of the NV centers and inhomogeneously by the inhomogeneity of the external magnetic field. In samples containing less NV centers, we can resolve this structure suggesting that the homogeneous broadening might by the more significant effect. It seems that in this NV center concentration regime and with this relatively small hyperfine splitting it is troublesome to give an estimation of the individual linewidth and hence $T_2$.

\clearpage
\section{Orientation dependence of ZFS-induced relaxation}
In the high-field approximation, where the quantization axis is along the external magnetic field, the zero field splitting Hamiltonian is given by
\begin{equation} 
\hat{H}_{\text{ZFS}}=\frac{1}{2}(3\ \textnormal{cos}^2\theta -1) D \hat{S}_z^2,
\end{equation}
where $\theta$ is the angle between the external magnetic field and the unique axis of the ZFS tensor. In general, the magnitude of the ZFS at an angle \textit{$\theta$} with respect to the unique axis of the tensor is proportional to $f(\theta)=3\ \textnormal{cos}^2\theta -1$. The relaxation rate in the Redfield regime is proportional to the square of the amplitude of the fluctuations hence to $f^2(\theta)$. 
In our pulse EPR measurements, the relaxation time of NV centers is acquired in two different orientations with respect to the external magnetic field. To give an estimation of the ratio of the longitudinal relaxation times one has to calculate the ratio of the above defined $f^2(\theta)$.
In the first case, the NV centers are parallel to the external magnetic field therefore the modulating ZFS fluctuations are perpendicular to the quantization axis hence $f^2(\theta=90^\circ)=1$. In the second case, the NV axis includes the tetrahedral angle ($109.47^\circ$) with the magnetic field. By applying simple trigonometric identities 
\begin{equation} 
\textnormal{cos}(\theta -90°)=\textnormal{sin}(\theta)=\sqrt{1-\textnormal{cos}^2\theta}
\end{equation}
and that the cosine of the tetrahedral angle is explicitly 1/3, we get $f(\theta=109.47^\circ)=5/3$, and $f^2(\theta=109.47^\circ)=25/9$.
Based on this argument, the ratio of ZFS induced relaxation for the two given orientations should be $25/9\approx 2.77$.
\clearpage
\section{Comparison of CW EPR spectrum to field swept electron spin echo}
\begin{figure}[!h]
\centering
\includegraphics*[width=1\linewidth]{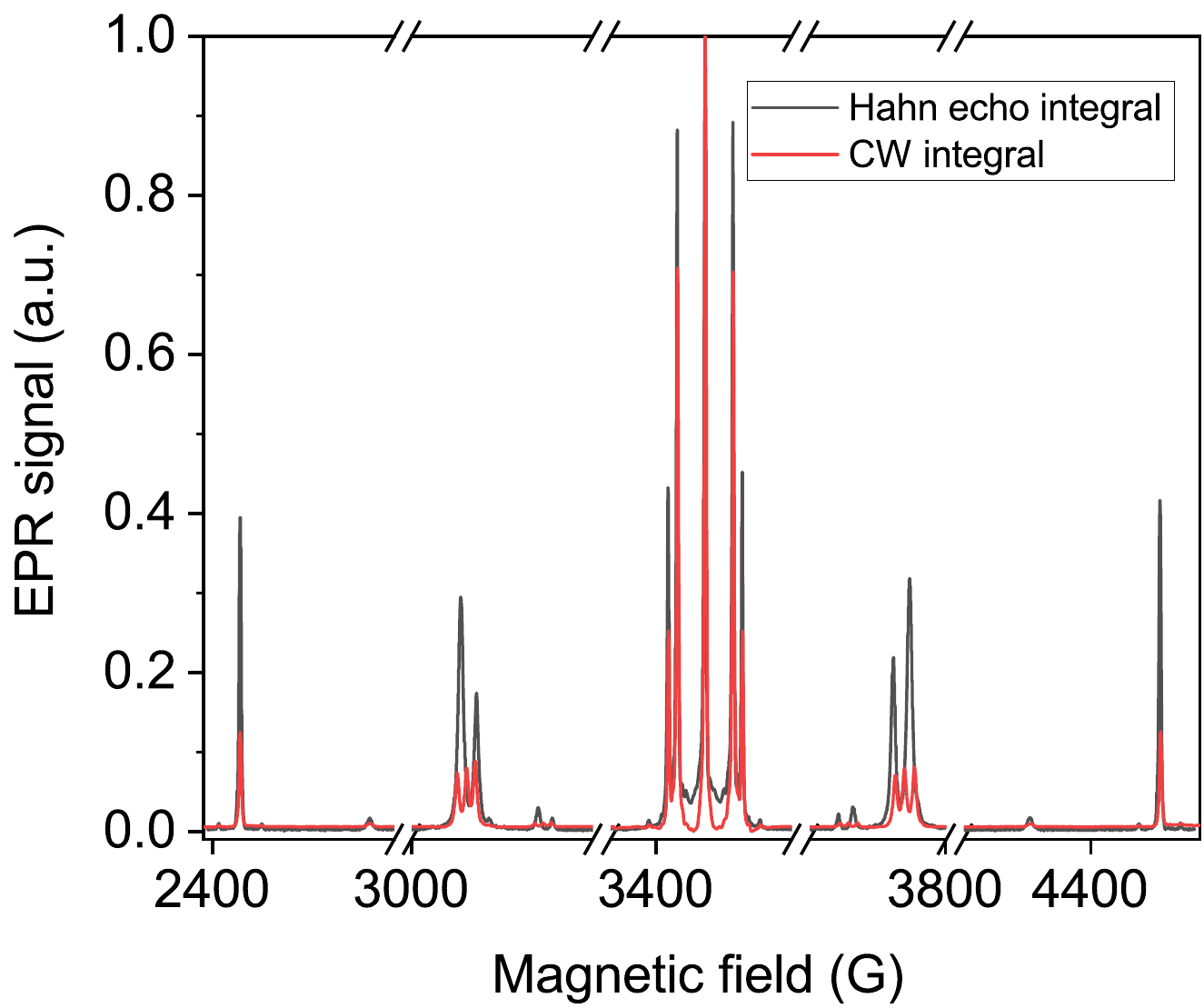}
\caption{\emph{CW EPR and field swept Hahn-echo spectrum of an ensemble of NV centers. Both spectra are normalized to the $m=0$ P1 line. }}
\label{cw_vs_echo}
\end{figure}

Fig. \ref{cw_vs_echo} shows continuous-wave EPR spectrum and field swept Hahn-echo spectrum both of them measured on sample S1. As the integral of the echo is proportional to the relaxation time, the NV/N intensity ratio becomes bigger in the echo than in the CW spectrum. By normalizing both spectra to the central nitrogen (P1) line, this results in stronger NV resonances in the echo compared to the CW spectrum.
\clearpage

\section{FT-ESEEM}
\begin{figure}[!h]
\centering
\includegraphics*[width=1\linewidth]{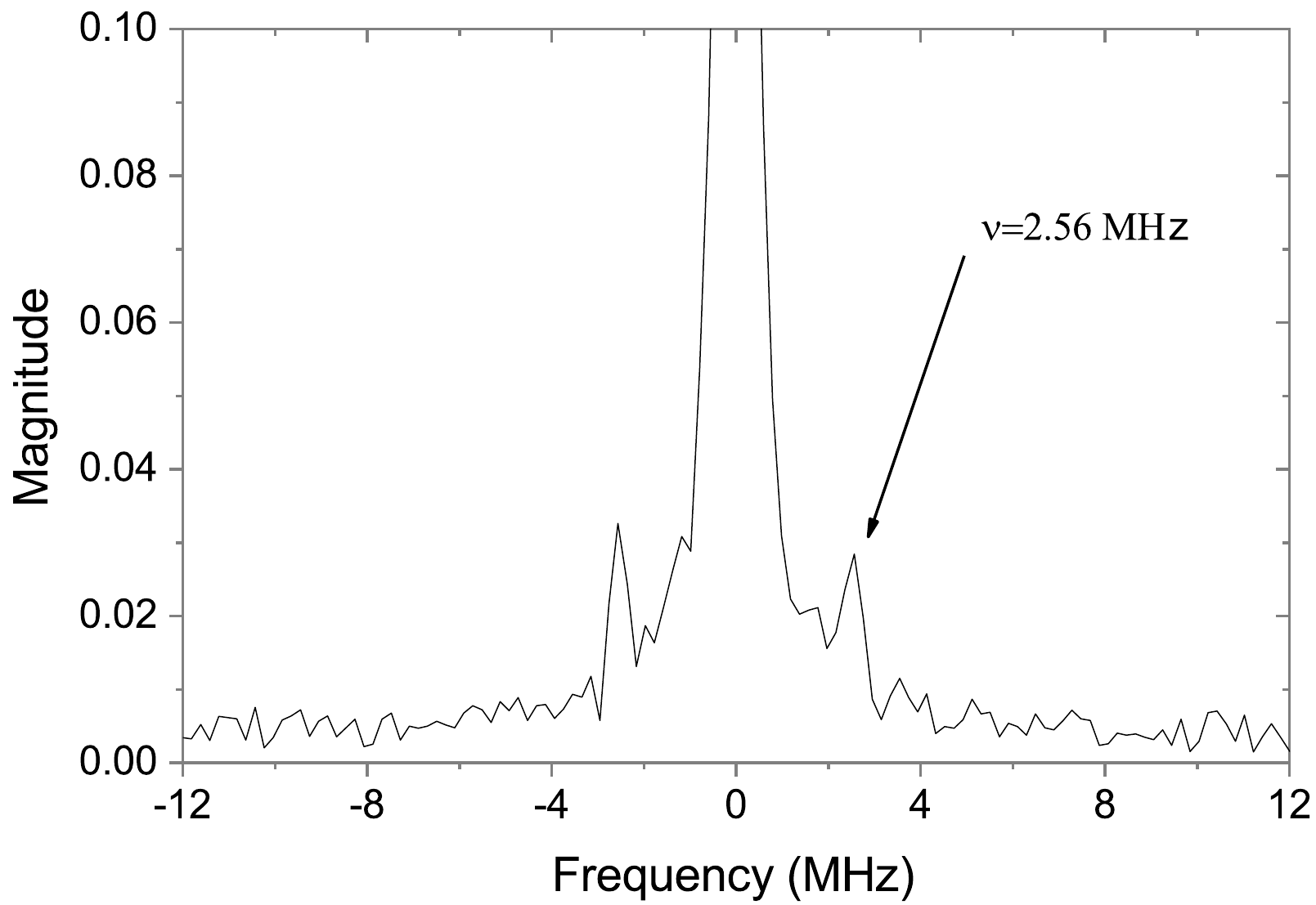}
\caption{\emph{FT-ESEEM of NV center in parallel magnetic field ($\text{B}=243\ \text{mT}$). The modulation is caused by distant $\ce{^{13}C}$ nuclei.}}
\label{SM5}
\end{figure}

Fig. \ref{SM5} shows the Fourier transform of the time domain electron spin envelope modulation (ESEEM) signal of NV centers aligned parallel to the external magnetic field (black curve in Fig. 4 in the main text). From the FT-ESEEM one peak can be identified at $\nu=2.56\ \text{MHz}$. The source of the modulation is that during the second microwave pulse some of the nuclear spin ($\ce{^{13}C}$ in this particular case) flip. The magnetic field experienced by electron spins hence their Larmor frequency changes. This causes the different spin packets to reach the maximum echo at slightly different times. The time domain ESEEM signal is the echo amplitude measured at time $\tau$ after the second pulse. As the spin packets precess with different frequencies, one has two echoes that are not in phase. By changing $\tau$, the accumulated phase difference changes hence the measured echo amplitude oscillates with the frequency difference of the spin packets. Taking the Fourier transform as a function of the pulse separation time $\tau$, the measured frequency is $\nu=2.56\ \text{MHz}$ which is close to the nuclear resonance frequency of $\ce{^{13}C}$ at this given magnetic field ($B=243\ \text{mT}$) implying that the modulation is due to the weakly coupled distant $\ce{^{13}C}$ nuclei. 

\clearpage
\section{ENDOR data}
Here we list the transition frequencies indicated on the ENDOR spectra in Fig. 5 in the main text. All values are given in MHz units.
\begin{center}
\begin{tabular}{| c | c | c | c | c | c | c | c |}
\hline
$f_5$ & $f_8$ & $\ast$ & $f_7$ & $f_6$ & ? & ? & ?\\
\hline
3.565 & 4.144 & 4.778 & 5.714 & 6.303 & 6.920 & 7.301 & 8.880\\   
\hline
\hline
$f_1$ & $\ast$ & $f_3$ & $f_4$ & $f_2$ & ? & &  \\
\hline
2.034 & 2.605 & 4.194 & 5.684 & 7.840 & 9.085 & &  \\ 
\hline
\end{tabular}
\end{center}

\end{document}